\documentstyle{article}
\input epsf.tex
\def\.{\!\cdot\!}
\def\:{\cdots}
\def\[{\left[}
\def\]{\right]}
\def\({\left(}
\def\){\right)}

\def\bi{\begin{itemize}}

\def\ei{\end{itemize}}
\def\be{\begin{eqnarray}}
\def\ee{\end{eqnarray}}
\def\bn{\begin{enumerate}}
\def\en{\end{enumerate}}
\def\h{{1\over 2}}

\def\nn{\nonumber\\}

\def\r2{\sqrt{2}}

\begin{document}
\title{
Time-Ordered Products and Exponentials\cite{ezawa}}
\author{C.S. Lam\cite{email}}
\date{Department of Physics, McGill University\\
3600 University St., Montreal, QC, Canada H3A 2T8}
\maketitle

\begin{abstract}
\noindent I discuss a formula decomposing
the integral of time-ordered
products of operators into sums of products of 
integrals of time-ordered commutators.  
The resulting factorization enables summation of
an infinite series
to be carried out to yield an explicit formula for the 
time-ordered exponentials. 
The Campbell-Baker-Hausdorff formula and the nonabelian eikonal formula
obtained previously are both special cases of this result.
\end{abstract}

\section{Introduction}

It is a great pleasure to dedicate this article to
Hiroshi Ezawa on the occasion of his 65th birthday. 
I am priviledged to have known him for 35 years, 
and am proud to see that he has such an illustrious career.
I wish him and his wife Yoshiko 
the very best, healthy and a long life that will last for
 at least another 35 years. 

True to the theme of this Workshop I will talk about
a mathematical method that has applications in quantum mechanics.
This is a method that could simplify
computations of (integrals of) 
time-ordered products and time-ordered exponentials $U$.
These quantities are ubiquitous in quantum mechanics. 
In one form or another they describe the time-evolution opreator of 
quantum systems, the non-integrable phase factor (Wilson
line) of Yang-Mills theories, and the
scattering amplitudes of perturbation diagrams. 
The method relies on a {\it decomposition formula}, which
expresses $U$ in terms of more primitive
quantities $C$, the time-ordered commutators. 

I shall 
concentrate in what follows to the description of this 
and other related formulas.
There is certainly no time for the proof and very little to illustrate
the applications. For those I refer the readers to the literature
\cite{LL1,LL2,FHL,FL,L}.

\section{Preliminaries}

Let $H_i(t)$ be operator-valued functions of time.
No attempts will be made to discuss domains and convergence problems
of these operators.
Let $[s]=[s_1s_2\cdots s_n]$ be a permutation of the $n$ numbers $[12\cdots n]$, and $S_n$ the corresponding permutation group. We
define the 
{\it time-ordered product} $U[s]$ to be the integral
\be
U[s]=U[s_1s_2\cdots s_n]=\int_{R[s]}dt_1dt_2\cdots
 dt_nH_{s_1}(t_{s_1})H_{s_2}(t_{s_2})\cdots H_{s_n}(t_{s_n})\label{Us}\ee
taken over the hyper-triangular region $R[s]=\{T\ge t_{s_1}\ge t_{s_2}
\ge\cdots\ge t_{s_n}\ge T'\}$, with operator $H_{s_i}(t_{s_i})$
standing to the left of $H_{s_j}(t_{s_j})$ if $t_{s_i}>t_{s_j}$.
The  average of $U[s]$ over all permutations $s\in S_n$
will be denoted by $U_n$:
\be
U_n&=&{1\over n!}\sum_{s\in S_n}U[s]\label{Un}.\ee

The  {\it decomposition formula}
expresses $U_n$ in terms of the
{\it time-ordered commutators} $C[s]=C[s_1s_2\cdots s_n]$.
These are defined
analogous to $U[s_1s_2\cdots s_n]$, but with the products of $H_i$'s
replaced by their nested multiple commutators:
\be
C[s]=\int_{R[s]}&&dt_1dt_2\cdots
 dt_n\nn
&&[H_{s_1}(t_{s_1}),[H_{s_2}(t_{s_2}),[\cdots ,[H_{s_{n-1}}(t_{s_{n-1}}),
H_{s_n}(t_{s_n})]\cdots]]].\label{Cs}\ee
For $n=1$, we define $C[s_i]=U[s_i]$.
Similarly, the operator $C_n$ is defined to be the average 
of $C[s]$ over all permutations $s\in S_n$:
\be
C_n&=&{1\over n!}\sum_{s\in S_n}C[s]\label{Cn}.\ee

It is convenient to use a `{\it cut}' (a vertical bar) 
to denote products of $C[\cdots]$'s.
For example, $C[31|2]\equiv C[31]C[2]$, and $C[71|564|2|3]\equiv
C[71]C[564]
C[2]C[3]$. Given a sequence $[s]$ of numbers with $s\in S_n$,
its {\it cut sequence} $[s]_c$ is obtained from
$[s]$ by inserting cuts at the appropriate places. A cut
is inserted after $s_i$ iff $s_i<s_j$ for all $i<j$.
In other words, we should proceed from
left to right and put a cut after the smallest number encountered.
The first cut is therefore put after the number `1'; the second
after the smallest number to the right of `1', etc.
For example, $[5413267]_c=[5431|2|6|7]$, and
$[1267453]_c=[1|2|67453]$.

\section{General Decomposition Formula}
The main formula \cite{L} states that
\be
n!U_n=\sum_{s\in S_n}U[s]=\sum_{s\in S_n}C[s]_c.
\label{thm}
\ee

For illustrative purposes 
here are explicit formulas for $n=1,2,3$, and 4:
\be
1!U_1&=&C[1]\nn
2!U_2&=&C[1|2]+C[21]\nn
3!U_3&=&C[1|2|3]+C[21|3]+C[31|2]+C[1|32]+C[321]+C[231]\nn
4!U_4&=&C[1|2|3|4]+C[321|4]+C[231|4]+C[421|3]+C[241|3]+C[431|2]\nn
&+&C[341|2]
+C[1|432]+C[1|342]+C[21|43]+C[31|42]+C[41|32]\nn
&+&C[21|3|4]
+C[31|2|4]+C[41|2|3]+C[1|32|4]+C[1|42|3]+C[1|2|43]\nn
&+&
C[4321]+C[3421]+C[4231]+C[3241]+C[2341]+C[2431]
\label{ex1}\ee
Using a filled circle with $n$ lines on top to indicate $n!U_n$,
and an open circle for $C[s]$, these formulas can be expressed
graphically as shown in Fig.~1.

\begin{figure}
\vskip -8cm
\centerline{\epsfxsize 4.7 truein \epsfbox {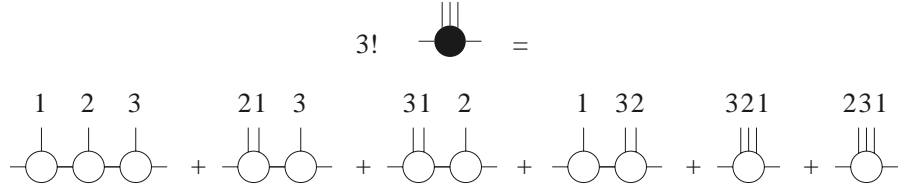}}
\nobreak
\vskip -1cm\nobreak
\caption{The decomposition of $3!U_3$ in terms of $C[s]$'s.}
\end{figure}

\section{Special Decomposition Formula}

Great simplification occurs when all $H_i(t)=H(t)$ are identical, 
for then $U[s]$ and $C[s]$ depend only on $n$ 
but not on the particular
$s\in S_n$. In that case all $U[s]=U_n$
and all $C[s]=C_n$, and the decomposition theorem becomes
\cite{L}
\be
U_n&=&\sum\xi(m_1m_2\cdots m_k)C_{m_1}C_{m_2}\cdots C_{m_k},\nn
\xi(m_1m_2\cdots m_k)&=&\prod_{i=1}^k\[\sum_{j=i}^km_j\]^{-1}\label{cor1}\ee
The sum in the first equation is taken over all $k$, and 
all $m_i>0$ such that $\sum_{i=1}^km_i=n$.
The quantity $\xi(m_1\cdots m_k)^{-1}$ is just the product of the number of numbers
to the right of every cut (times $n$).
Note that  it is 
{\it not} symmetric under the interchange of the $m_i$'s.
It is this asymmetry that makes the formulas for $K_n$
in (\ref{exp}) rather complicated.

We list below this special decompositions up to $n=5$:
\be
1!U_1&=&C_1\nn
2!U_2&=&C_1^2+C_2\nn
3!U_3&=&C_1^3+2C_2C_1+C_1C_2+2C_3\nn
4!U_4&=&C_1^4+6C_3C_1+2C_1C_3+3C_2^2+3C_2C_1^2+2C_1C_2C_1+C_1^2C_2
+6C_4\nn
5!U_5&=&C_1^5+24C_4C_1+6C_1C_4+12C_3C_2+8C_2C_3+12C_3C_1^2+6C_1C_3C_1\nn
&&+2C_1^2C_3+8C_2^2C_1+4C_2C_1C_2+3C_1C_2^2+4C_2C_1^3+3C_1C_2C_1^2\nn
&&+2C_1^2C_2C_1+C_1^3C_2.
\label{ex2}
\ee
The graphical expression for $U_3$ is given in Fig.~2.

\begin{figure}
\vskip -8cm
\centerline{\epsfxsize 4.7 truein \epsfbox {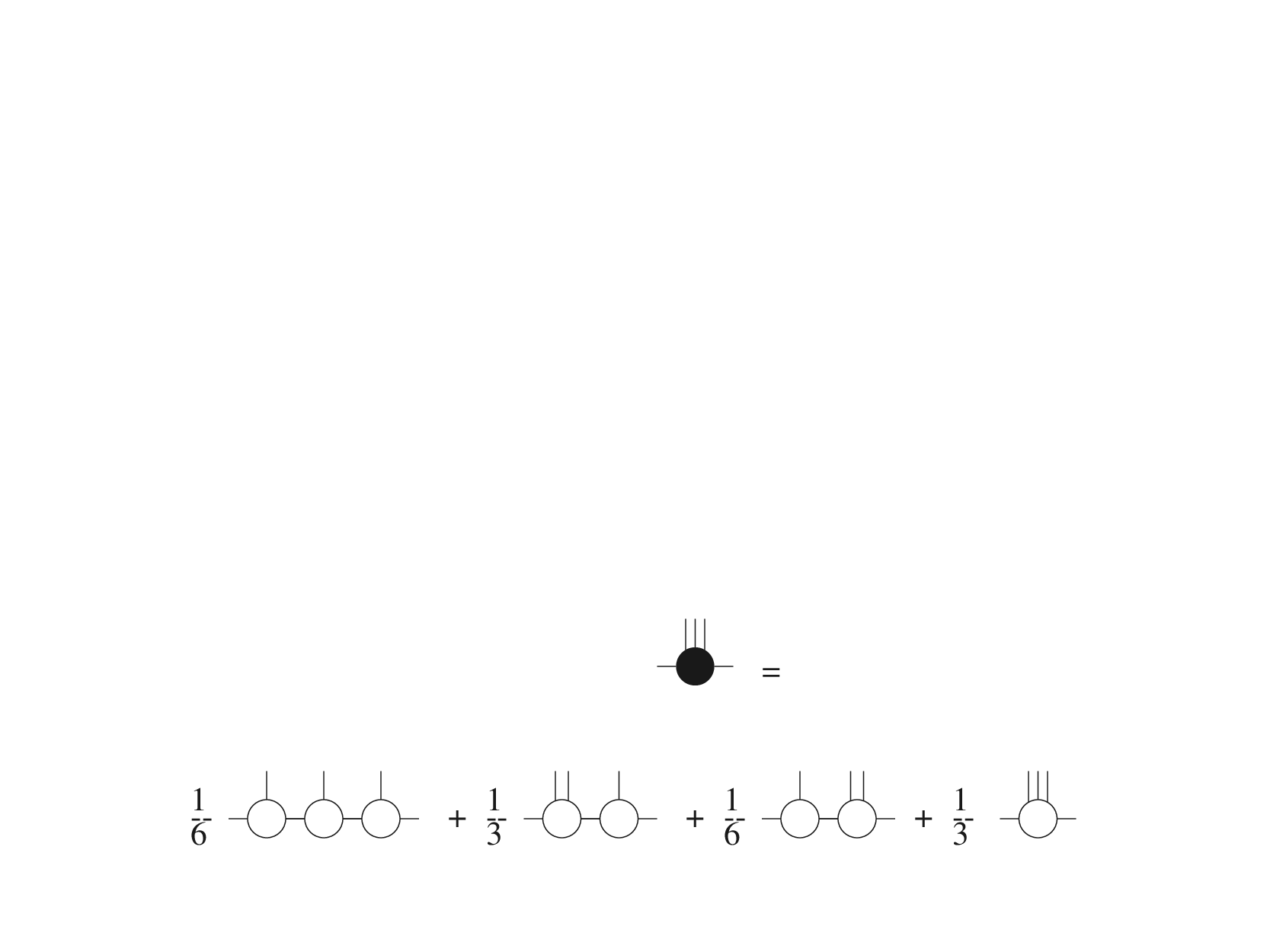}}
\nobreak
\vskip -1cm\nobreak
\caption{The decomposition of $U_3$ in terms of $C_m$'s.}
\end{figure}

\section{Exponential Formula for 
Time-Ordered Exponentials} 

The time-ordered exponential 
\be
{\cal U}&=&T\(\exp\(\int_{T'}^TH(t)dt\)\]
=\sum_{n=0}^\infty U_n\label{toexp}
\ee
can be computed from the time-ordered products
$U_n$. 
The factorization character in (\ref{cor1}) 
and (\ref{ex2}) suggests that
it may be possible to sum up the power series $U_n$
to yield an explicit exponential function of the $C_n$'s.
This is indeed the case.

\subsection{Commutative $C_i$'s}

First assume all the $C_{m_i}$ in (\ref{cor1})
commute with one another. Then it is possible to show that
\cite{L} 
\be
{\cal U}&=&\sum_{n=0}^\infty {U_n}\nn
&=&\prod_{j=1}^\infty \sum_{m=0}^\infty {1
\over j^{m}m!}C_j^{m}\nn
&=&\exp\[\sum_{j=1}^\infty{C_j\over j}\].\label{symexp}\ee
\subsection{General Exponential Formula}

In general the $C_j$'s do not commute with one another so
the exponent in (\ref{symexp}) must be corrected by terms
involving commutators of the $C_j$'s. The exponent $K$
can be computed by taking the logarithm of $U$ \cite{L}:
\be
U&=&1+\sum_{n=1}^\infty U_n=\exp[K]\equiv
\exp\[\sum_{i=1}^\infty K_i\],\nn
K&=&\ln\[1+\sum_{n=1}^\infty U_n\]=
\sum_{\ell=1}^\infty {(-)^{\ell-1}\over\ell}\[\sum_{n=1}^\infty
U_n\]^\ell.\label{log}\ee
The resulting expression must be expressible
 as (multiple-)commutators of the 
$C$'s. In other words, only commutators of $H(t)$, 
in the form of $C_m$ and their commutators, may enter into $K$.
This is so because in the special case when $H(t)$ is a member of
a Lie algebra, ${\cal U}$ is a member of the corresponding Lie group
and so $K$ must also be a member of a Lie algebra.

By definition, $K_i$ contains $i$ factors of $H(t)$. 
Calculation for the first five gives \cite{L}
\be
K_1&=&\ \ C_1\nn
K_2&=&{1\over 2}C_2\nn
K_3&=&{1\over 3}C_3+{1\over 12}[C_2,C_1]\nn
K_4&=&{1\over 4}C_4+{1\over 12}[C_3,C_1]\nn
K_5&=&{1\over 5}C_5+{3\over 40}[C_4,C_1]+{1\over 60}[C_3,C_2]
+{1\over 360}[C_1,[C_1,C_3]]+\nn
&&\quad +{1\over 240}[C_2,[C_2,C_1]]
+{1\over 720}[C_1,[C_1,[C_1,C_2]]]\label{exp}\ee

$K_n$ consists of $C_n/n$, plus the compensating terms in the form of commutators
of the $C$'s. By counting powers of $H(t)$ it is clear that
 the subscripts of
these $C$'s must add up to $n$, but beyond that all independent
commutators and multiple commutators may appear.
For that reason it is rather difficult to obtain an explicit
fomula valid for all $K_n$,
if for no other reason than the fact that new 
commutator structures appear at
every new $n$. It is however very easy to 
compute $K_n$ using (\ref{log}) with the help of a computer.
This is actually how $K_5$ was obtained.

Moreover, when we stick to commutators of a definite structure,
their coefficients in $K$ can be computed. For example , the coefficient
of the commutator term $[C_m,C_n]$ in any $K_{m+n}$ can be shown to be
\be
\eta_2={n-m\over 2mn(m+n)}.\ee
See Ref.~\cite{L} for similar formulas for multiple commutators.

\section{Applications}

These formulas can be applied to mathematics and physics in various
ways, depending on our choice of $H_i(t)$ and the integration interval
$[T',T]$. If we choose the interval to be $[T',T]=[0,2]$, 
and the operator $H(t)$ to be $P$ for $t\in[1,2]$ and $Q$ for $t\in[0,1]$,
then ${\cal U}=\exp(P)\exp(Q)$, $C_1=P+Q$, $C_{m+1}=(ad\ P)^m\. Q/m!$,
and eqs.~(\ref{log}) and (\ref{exp}) lead to \cite{L} the 
{\it Baker-Campbell-Hausdorff} formula
\be
\exp(P)\.\exp(Q)&=&\exp[K_1+K_2+K_3+K_4+K_5+\cdots]\nn
K_1&=&P+Q\nn
K_2&=&\h[P,Q]\nn
K_3&=&{1\over 12}[P,[P,Q]]+{1\over 12}[Q,[Q,P]]\nn
K_4&=&-{1\over 24}[P,[Q,[P,Q]]]\nn
K_5&=&-{1\over 720}[P,[P,[P,[P,Q]]]]-{1\over 720}[Q,[Q,[Q,[Q,P]]]]\nn
&&+{1\over 360}[P,[Q,[Q,[Q,P]]]]+{1\over 360}[Q,[P,[P,[P,Q]]]]\nn
&&+{1\over 120}[P,[Q,[P,[Q,P]]]]+{1\over 120}[Q,[P,[Q,[P,Q]]]],
\label{BH}\ee
The case
when $[P,Q]$ commutes with $P$ and $Q$ is well known. 
In that case all $K_n$ for $n\ge 3$ are zero.
Otherwise,
up to and including $K_4$ this formula 
can be found in eq.~(15), \S 6.4, Chapter II of Ref.~\cite{bourbaki}.

By choosing the interval to be $[T',T]=[-\infty,\infty]$,
and the operators to be $H(t)=\exp(ip\. k_it)V_i$,
we obtain a {\it nonabelian eikonal formula} 
\cite{LL1,FHL,L} useful in
physics. In that case $U_n$ is the $n$th order tree
amplitude for an energetic particle with four-momentum
$p^\mu$ to emit  $n$ bosons with momenta $k_i^\mu\ll p^0$
via vertex factors $V_i$. The decomposition formula
(\ref{thm}) can then be interpreted as a repackaging
of the tree amplitude into terms in which 
interference patterns in the spacetime and
the internal quantum number variables are explicitly
displayed. Such (destructive) interferences lead to
cancellations, and the formula can be conveniently used
to demonstrate the cancellations necessary for the 
self consistency of baryonic amplitudes in large-$N_c$ QCD
\cite{LL1,LL2}. It can also be used to 
obtain a simple understanding as to why
gluons reggeize in QCD but photons do not reggeize in QED
\cite{FHL,FL}.


\begin{thebibliography}{9}
\bibitem[**]{ezawa} Contribution to the Second Jagna International Workshop
on Mathematical Methods of Quantum Physics, January 4-8, 1998, at
Jagna, Bohol, Philippines,
in honour of Prof. Hiroshi
Ezawa on the occasion of his 65th birthday.
\bibitem[*]{email} Email: Lam@physics.mcgill.ca
\bibitem{LL1}
C.S. Lam and K.F. Liu, {\it Nucl. Phys.} {\bf B483} (1997) 514.
\bibitem{LL2} C.S. Lam and K.F. Liu, {\it Phys. Rev. Lett.}
{\bf 79} (1997) 597. 
\bibitem{FHL}Y.J. Feng, O. Hamidi-Ravari, and C.S. Lam,
{\it Phys. Rev. D} {\bf 54} (1996) 3114.
\bibitem{FL} Y.J. Feng and C.S. Lam, {\it Phys. Rev. D}
{\bf 55} (1997) 4016.
\bibitem{L}C.S. Lam, to be published.
\bibitem{bourbaki} N. Bourbaki, {\it Lie Groups and Lie Algebras},
(Hermann, 1975).

\end{thebibliography}
\end{document}